\documentclass[sigconf,nonacm]{acmart}

\usepackage{amsmath,amsfonts,amsthm}
\usepackage{algorithm,algorithmic}

\usepackage{xspace}
\usepackage{graphicx}
\usepackage{textcomp}
\usepackage{xcolor}
\usepackage{listings}
\usepackage{multirow}
\usepackage{pifont}
\usepackage{hyperref}
\usepackage{pifont}
\usepackage{enumitem}
\usepackage{url}
\usepackage{booktabs}
\usepackage{tabularx}
\usepackage{multirow,array}
\usepackage{threeparttable}
\usepackage{caption}
\usepackage{subcaption}
\setcopyright{none}
\renewcommand\footnotetextcopyrightpermission[1]{}
\usepackage{tikz}

\def\BibTeX{{\rm B\kern-.05em{\sc i\kern-.025em b}\kern-.08em
    T\kern-.1667em\lower.7ex\hbox{E}\kern-.125emX}}

\newtheorem*{theorem*}{Theorem}

\usepackage{tcolorbox}
\definecolor{mycolor}{rgb}{0.122, 0.435, 0.698}
\definecolor{gray1}{gray}{0.3}

\definecolor{codegreen}{rgb}{0,0.6,0}
\definecolor{codegray}{rgb}{0.5,0.5,0.5}
\definecolor{codepurple}{rgb}{0.58,0,0.82}
\definecolor{backcolour}{rgb}{0.95,0.95,0.92}
\lstdefinestyle{mystyle}{
    commentstyle=\color{codegreen},
    keywordstyle=\color{magenta},
    numberstyle=\tiny\color{codegray},
    stringstyle=\color{codepurple},
    basicstyle=\tiny\ttfamily,
    breakatwhitespace=false,
    breaklines=true,
    captionpos=b,
    keepspaces=true,
    numbers=left,
    numbersep=5pt,
    showspaces=false,
    showstringspaces=false,
    showtabs=false,
    tabsize=2,
    columns=fixed
}
\newcommand{\result}[1]{%
\begin{tcolorbox}[colframe=mycolor,boxrule=0.5pt,arc=4pt,
      left=6pt,right=6pt,top=6pt,bottom=6pt,boxsep=0pt,width=\columnwidth]%
      {\emph{#1}}
\end{tcolorbox}%
}

\definecolor{darkgreen}{rgb}{0.0, 0.5, 0.0}
\definecolor{darkred}{rgb}{0.82, 0.1, 0.26}
\newcommand{\xmark}{\textcolor{darkred}{\ding{55}}\ }%

\newcommand{\afc}{\textsc{AFLChaos}\xspace}
\newcommand{\aflnet}{\textsc{AFLNet}\xspace}

\AtBeginDocument{%
  \providecommand\BibTeX{{%
    Bib\TeX}}}

\begin{document}

\title{Finding Bug-Inducing Program Environments}

\author{Zahra Mirzamomen}
\email{zahra.mirzamomen@monash.edu}
\affiliation{%
  \institution{Monash University}
  \streetaddress{}
  \city{}
  \state{}
  \country{}
  \postcode{}
}

\author{Marcel B{\"o}hme}
\authornotemark[1]
\email{marcel.boehme@acm.org}
\affiliation{%
  \institution{Max Planck Institute for Security and Privacy}
  \streetaddress{}
  \city{}
  \state{}
  \country{}
  \postcode{}
}


\begin{abstract}
  Some bugs cannot be exposed by program \emph{inputs}, but only by certain program \emph{environments}. During execution, most programs access various resources, like databases, files, or devices, that are external to the program and thus part of the program's environment.
In this paper, we present a coverage-guided, mutation-based environment synthesis approach of bug-inducing program environments.
Specifically, we observe that programs interact with their environment via dedicated system calls and propose to intercept these system calls (i)~to \emph{capture} the  resources accessed during the first execution of an input as initial program environment, and (ii)~\emph{mutate} copies of these resources during subsequent executions of that input to generate slightly changed program environments. Any generated environment that is observed to increase coverage is added to the corpus of environment seeds and becomes subject to further fuzzing. Bug-inducing program environments are reported to the user.

Experiments demonstrate the effectiveness of our approach. We implemented a prototype called \textsc{AFLChaos} which found bugs in the resource-handling code of five (5) of the seven (7) open source projects in our benchmark set (incl. OpenSSL). Automatically, \textsc{AFLChaos} generated environments consisting of bug-inducing databases used for storing information, bug-inducing multimedia files used for streaming, bug-inducing cryptographic keys used for encryption, and bug-inducing configuration files used to configure the program. To support open science, we publish the experimental infrastructure, our tool, and all data.
\end{abstract}

\maketitle

\section{Introduction}\label{sec:intro}
Recently, coverage-guided greybox fuzzing has become one of the most suc\-cess\-ful automatic bug finding techniques in practice. For instance, in the past five years, the OSSFuzz and ClusterFuzz projects have found 29{,}000+ bugs in products like Chrome and 40{,}000+ bugs in over 650 open source projects \cite{clusterfuzz,ossfuzz2}. Given a corpus of inputs, a greybox fuzzer generates new inputs by mutation and adds every coverage-increasing input to the seed corpus for further fuzzing.

\begin{figure}
    \centering
    \includegraphics[trim={2.5cm 7.4cm 3cm 0.9cm},clip, width=\columnwidth]{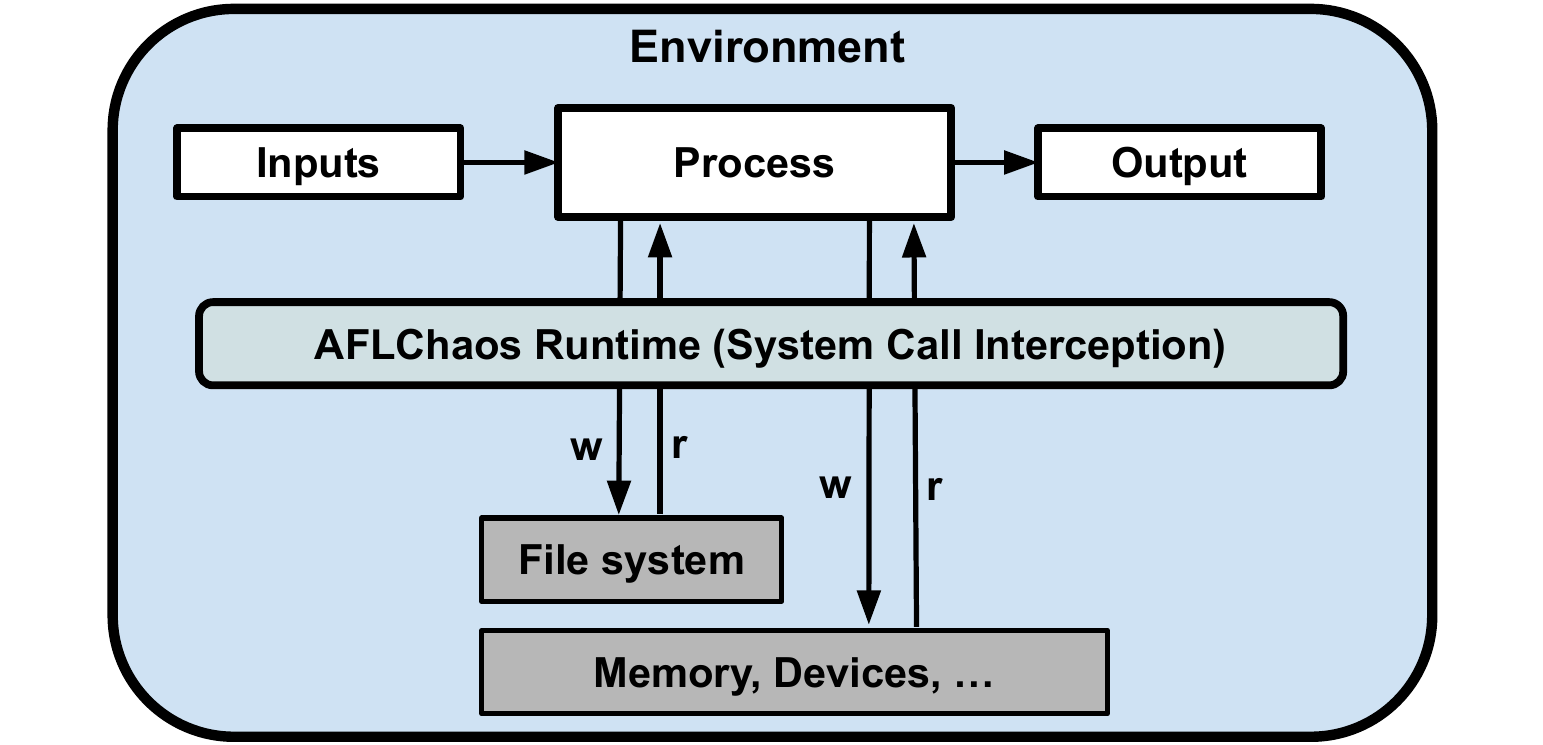}
    \caption{A program being executed within an environment.\vspace{-0.2cm}}
    \label{fig:environm}
\end{figure}

However, program behavior not only depends on its~input, but also the environment within which it operates (Fig.~\ref{fig:environm}). 
If an unprivileged user controls the environment of a privileged process and exploits an environment-dependent bug, they can launch privilege escalation attacks~\cite{pescalate2003}.
To discover such bugs, the program must be executed in a bug-inducing environment.

In this paper, we introduce \afc, a technique that allows us to generate such bug-inducing  environments. Towards this objective, there are several challenges that we need to overcome. Challenge~1 is that the bug-inducing environment must be \emph{realistic}: A user should be able to realize this environment on an actual machine without interrupting other processes. Challenge~2 is that the generated environments must actually \emph{impact program behavior}: The program under test likely only interacts with a small part of the environment.

To address Challenge~1 and generate realistic environments, \afc is inspired by key ingredients of coverage-guided greybox fuzzing. Firstly, we use a \emph{mutation-based approach}. Instead of generating each environment from scratch, we consider the current program environment as a seed environment which is  corrupted only slightly by mutation to generate other realistic environments. Secondly, we use a \emph{coverage-guided approach}, adding every coverage-increasing program environment to the corpus of environments. However, if we took a snapshot-based approach to capture, mutate, and restore the \emph{entire} state of a virtual machine \cite{Nyx2022,nyxnet}, many of the mutations would be redundant and not actually impact the program behavior. So, to address Challenge~2, we capture, mutate, and restore the \emph{interactions with the environment}.

We observe that programs access environment resources via dedicated system calls. We propose, during the execution of a seed input, to intercept each system call and make a copy of the accessed resource or the returned data.\footnote{As an added advantage, \afc tackles flakiness due to external state. \afc restores the environment before every execution of the program. For instance, \afc automatically restores databases, reverts changes, and clears persistent cache.} During subsequent executions of the program, we propose to return mutated copies instead of the data from the original resource. For every environment resource, \afc maintains a separate resource-specific seed corpus. The sequence of resource copies that is used during an execution of a program is known as program environment. Generated environments are stored in a separate corpus if they are coverage-increasing.

\begin{figure*}
  \centering
  \includegraphics[trim={0 10.4cm 0 0.2cm},clip,width=0.9\textwidth]{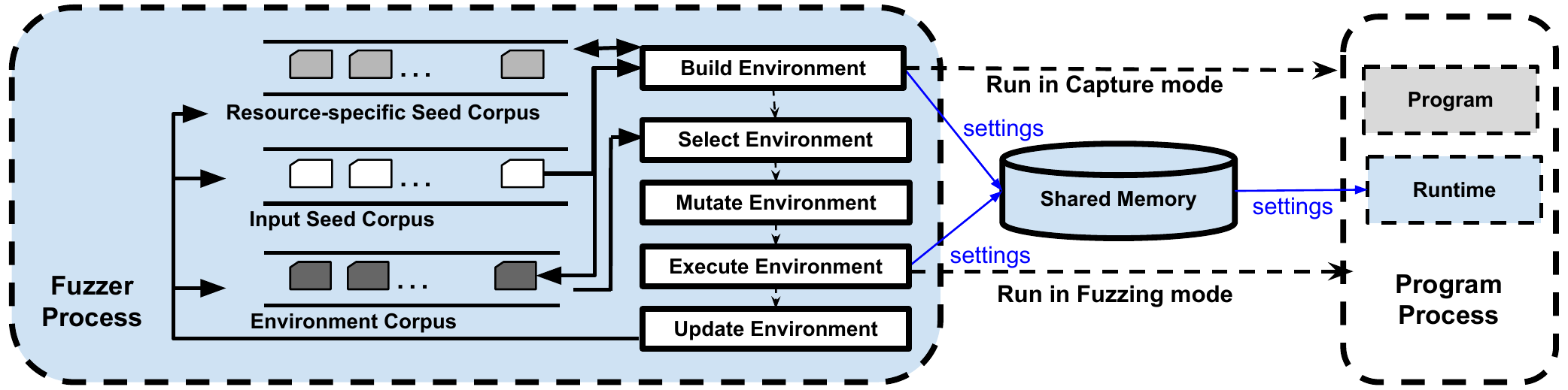}\vspace{-0.2cm}
  \caption{The overall architecture of \afc\vspace{-0.2cm}}
  \label{fig:arch}
\end{figure*}
We implemented \afc into the \aflnet network-enabled greybox fuzzer \cite{aflnet} to evaluate our coverage-guided, mutation-based environment synthesis on the coverage-instrumented program binaries of seven servers which implement different network protocols. However, we note that our approach is \emph{not} specific to fuzzing network protocols and can be implemented in other fuzzers. 

To investigate our main hypothesis that fuzzing the program environment in addition to the program input improves fuzzing effectiveness, we compared \afc against its baseline in experiments involving seven servers implementing different internet protocols (encryption, streaming, file sharing, messaging, VoIP) from the ProFuzzbench fuzzer benchmark \cite{profuzzbench}.\footnote{While other benchmarks only contain smaller system components, such as libraries \cite{fuzzbench,magma,fts}, we wanted to fuzz whole systems that, like the most widely-used systems, actually do interact with the environment. For instance, servers read/write certificates, databases, configurations, et cetera.} We found that \afc achieves 7\% more branch coverage than \aflnet. This increase is directly associated with new coverage of environment-handling code.

\afc found 13 previously undiscovered bugs in five of the seven (5/7) server programs---none of which \aflnet could find. A qualitative analysis revealed that these bugs can only be induced by specific program environments rather than server inputs (i.e., user requests). We found that the root causes of those environment-induced bugs are located in source code that is handling configuration files, databases, certificates and keys, as well as the multimedia files intended for streaming. If a malicious, unprivileged user can modify these environment resources of the corresponding privileged process, they may be able to launch a privilege escalation attack. 

In terms of performance overhead, for four of seven (4/7) servers, we found that fuzzing the environment incurs substantial extra cost, often due to the changes rather than the runtime overhead. However, given the increased effectiveness, we find this additional cost appropriate. For the remaining three (3/7) servers, we found that fuzzing environment resources actually lead to a better performance because the program would exit faster for executions with invalid resources.

\vspace{0.1cm}
\noindent
In summary, this paper makes the following contributions.
\begin{itemize}
  \item We propose an environment-extended fuzzing framework that effectively finds bug-inducing program environments.
  \item We propose a novel approach in capturing, mutating and restoring solely the pertinent components of the environment by taking into consideration the real interactions of the program with the environment.
  \item We achieved 7\% more branch coverage than \aflnet, directly associated with the environment-handling code and found 13 bugs in five server programs, all induced by specific program environments rather than server inputs.
\end{itemize}
\textbf{Open Science}. To facilitate reproducibility, we publish our research prototype, experimental setup, data, and analysis.

{%
\centering
\url{https://anonymous.4open.science/r/AFLChaos\_Artifact-024B}
}

\section{\afc Architecture}\label{sec:arch}
\afc, our environment-extended fuzzing framework consists of two components: a \emph{runtime} which intercepts system calls by the program to access environment resources (i.e., the file system), and the \emph{fuzzer} which (i)~captures the resources accessed by the program when processing an input as environment seed, (ii)~mutates environment seeds to generate alternative environments, and (iii)~adds coverage-increasing environments as new environment seeds for further fuzzing.

\autoref{fig:arch} shows the general architecture of \afc. The \emph{runtime} resides in the same process as the program under test. A shared memory enables the inter-process communication  between the fuzzer process and the runtime (cf. \autoref{sec:runtime}). The fuzzer maintains three kinds of seed corpora (cf. \autoref{sec:fuzzer}), which are read and written to by the main building blocks of the fuzzing algorithm (cf. \autoref{sec:mainalg} \& Algs. \ref{alg:mutate}-\ref{alg:buildenvs}).

\subsection{Runtime: Capturing Environment Interactions}\label{sec:runtime}
\textbf{Preloading}. To capture the interaction of a program with its environment, most of the program's resource-related system calls are intercepted. Our approach works with any program binary by \emph{preloading} a shared library that implements our runtime. All major operating systems support preloading. In Unix-based operating systems, system calls can be intercepted using the \texttt{LD\_\-PRELOAD} environment variable which can be pointed to the location of our shared library. On MacOS, the \texttt{DYLD\_\-INSERT\_\-LIBRARIES} environment variable needs to be set. On Windows, a shared library can be injected into a binary using DLL injection (e.g., using \texttt{AppInit\_\-DLLs}).

\textbf{System resources}. In this work, we focus particularly on \emph{system resources} as part of the program environment.
\begin{itemize}
\item Firstly, system resources often have unique identifiers. This allows us to maintain resource-specific seed corpora, i.e., a set of variants for every (uniquely identified) resource that is accessed by a program during the execution of an input. In a classification of system calls according to their level of threat with respect to system penetration, Massi et al.~\cite{Massi2002} found such resource-related system calls to be within the most critical threat level. Vulnerabilities involving these system calls may allow full control of the system, e.g., via privilege escalation.
\item Secondly, access to resources is standardized in Unix operating systems: "Everything is a file" (unless it is a process) \cite{sanders2018unix}. This allows us to intercept only a few critical system calls related to the file system. The unique resource identifiers are called file descriptors.
\end{itemize}

Our runtime intercepts resource-related calls to the environment and returns mutated versions of those resources as instructed by the fuzzer. Given an input $t$ to be executed on the program, the runtime works in two modes (set by the fuzzer).
\begin{itemize}
\item In \emph{capture mode}, whenever a resource $r\in R$ is accessed during the execution of $t$, our runtime makes a copy of the \emph{original} resource $r$. Our runtime implementation stores a copy of resource $r$ as a file that carries the \emph{unique} resource identifier in the file name.\footnote{Later, the fuzzer will use those copies of the original resources as initial seeds for the resource-specific seed corpora.} Finally, the runtime returns to the program the original handle to the resource.
\item In \emph{fuzzing mode}, whenever a resource $r\in R$ is accessed during the execution of $t$, our runtime returns a handle to a potentially fuzzed copy of $r$. Exactly which copy of $r$ will be made available to the program is determined by the fuzzer.
\end{itemize}
\textbf{Runtime-Fuzzer-Program communication}.  As illustrated in \autoref{fig:arch}, the runtime and the program are executed within the same process. Intercepted system calls are transparently delegated to the runtime without any overhead. However, the runtime and fuzzer do not share the same process. So, we require a mechanism to facilitate inter-process communication. For maximum efficiency, we leverage a shared memory architecture.\footnote{This approach is inspired by AFL \cite{afl} which uses shared memory to collect coverage-feedback from the program's process.} Since our runtime process is killed after each execution of the program, we designed our runtime to be stateless. All state of \afc exists either in the fuzzer or the file system.

\subsection{Fuzzer: Modifying Environment Interactions}\label{sec:fuzzer}
The \afc fuzzer uses the runtime to capture and substitute the resource-related environment of a program in a coverage-guided manner. The fuzzer maintains three kinds of seed corpora and implements several novel mutation operators to support environment synthesis.

\textbf{Seeds}. \afc maintains resource-related and composite seeds in addition to traditional input seeds.
\begin{enumerate}
  \item \emph{Input seed corpus} $C_{\text{Inp}}$. Like other greybox fuzzers \cite{nyxnet,afl,aflnet,libfuzzer}, \afc maintains a corpus of coverage-increasing program inputs. 
  \item \emph{Resource-specific seed corpus} $C_r$. For each resource $r$ that is accessed during the execution of any program input $t\in C_\text{Inp}$, \afc maintains a resource-specific seed corpus $C_r$. During fuzzing, any generated coverage-increasing (mutated) copy of $r$ is added to $C_r$. The set of $C_r$ for all resources $r$ accessed during the execution of any input $t\in C_{\text{Inp}}$ is denoted by $\mathfrak{C}$.
  \item \emph{Environment corpus} $C_\text{Env}$. A configuration of program input and one (mutated) copy for each resource accessed during the execution of that input is called "environment". The initial environment $S(t,R)\in C_\text{Env}$ for an input $t\in C_\text{Inp}$ is constructed when executing $t$ with the runtime set to "capture mode" which records all resources $R$ accessed by $t$. Alternative environments $S'$ can be constructed by mutating an environment $S(t,R)\in C_\text{Env}$. Any generated coverage-increasing program environment is added to $C_\text{Env}$.
\end{enumerate}
Any crashing environment is stored separately for later triage and reproduction. In our experience, the well-defined structure of our generated environments simplifies the crash reproduction and allows us to pinpoint precisely which file is the crash-inducing environment resource.

\begin{algorithm}
\caption{Function \textsc{mutateEnviron}}
\label{alg:mutate}
\begin{algorithmic}
\REQUIRE{Environment seed $S=S(t,R)$}
\STATE \textbf{switch} \textsc{randomChoice}($1:3$)
      \begin{ALC@g}
      \STATE \textbf{case} 1:
        \begin{ALC@g}
          \STATE Resource copy $r=$ \textsc{randomChoice}($R$)
          \STATE New resource copy $r'=$ \textsc{fuzz}($r$)
          \STATE New set of resources $R' = (R \setminus \{r\})\cup \{r'\}$
          \STATE New environment $S'=S(t,R')$
          \STATE \textbf{break}
        \end{ALC@g}
      \STATE \textbf{case} 2:
        \begin{ALC@g}
          \STATE Resource copy $r=$ \textsc{randomChoice}($R$)
          \STATE New resource copy $r'=$ \textsc{randomChoice}($C_r$)
          \STATE New set of resources $R' = (R\setminus \{r\})\cup \{r'\}$
          \STATE New environment $S'=S(t,R')$
          \STATE \textbf{break}
        \end{ALC@g}
      \STATE \textbf{case} 3:
        \begin{ALC@g}
          \STATE New program input $t'=$ \textsc{randomChoice}($C_\text{Inp}$)
          \STATE New environment $S'=S(t',R)$
          \STATE \textbf{break}
        \end{ALC@g}
      \end{ALC@g}
    \STATE \textbf{end switch}
\ENSURE{Alternative environment $S'$}
\end{algorithmic}
\end{algorithm}

\textbf{Mutation operators}. As shown in Algorithm~\ref{alg:mutate},  we design three simple operators for an environment $S(t,R)\in C_S$. 
\begin{description}[leftmargin=1cm,itemsep=0.2cm]
  \item[OP\#1] \emph{Fuzz resource}. The first operator chooses a resource $r\in R$ at random, fuzzes the contents of the resource, and substitutes the chosen resource with the fuzzed version in the environment to generate an alternative environment $S(t, R')$. Since we store copies of a resource as files, we can reuse the existing mutation operators for file fuzzing to fuzz resources. Of course, this makes \afc amenable to the corresponding optimizations, such as structure-aware fuzzing \cite{aflsmart,nautilus}.
  \item[OP\#2] \emph{Switch resource}. The second operator chooses a resource $r\in R$ at random and replaces the copy in the environment with another resource-specific seed chosen at random from the resource-specific corpus $C_r$ to generate an alternative environment $S(t, R')$.
  \item[OP\#3] \emph{Switch input}. The third operator substitutes the input seed $t$ in the environment with another input seed $t'$ chosen at random from the input seed corpus $C_I$ to generate an alternative environment $S(t', R')$.
\end{description}
Coverage-increasing environments and relevant constituents (resources or inputs) are added to the corresponding corpora. Of course, depending on the application domain, one could imagine other mutation operators, such as structure-aware operators \cite{aflsmart,nautilus}, that could be implemented in Algorithm~\ref{alg:mutate}. Domain-specific mutation operators can be transparently implemented in the call to \textsc{fuzz}.

\textbf{Synchronization}.
We delegate the fuzzing of program inputs to a vanilla greybox fuzzer (here, \textsc{AFLNet}) that shares coverage-increasing program inputs with \afc via a synchronization step. Newly discovered program inputs contribute in construction of new environment seeds according to Algorithm~\ref{alg:mutate}. 

\begin{algorithm}\small
\caption{Coverage-guided Environment Synthesis}
\label{alg:fuzzing}
\begin{algorithmic}[1]
\REQUIRE {Initial input seeds $I$, Program $P$, Time budget \emph{timeout}}
\STATE {\emph{\textcolor{blue}{// Initializations}}}
\STATE {Input seed corpus $C_\text{Inp}=I$}\label{lst:line:init1}
\STATE {Resource-specific corpora $\mathfrak{C}=\emptyset$}
\STATE {Environment corpus $C_{\text{Env}}=\emptyset$}
\STATE {Crashing environments $C_{\text{\xmark}}=\emptyset$}\label{lst:line:inite}
\STATE {}
\STATE {\emph{\textcolor{blue}{// Record all resources accessed by initial inputs}}}
\STATE $(C_{\text{Inp}}$, $C_{\text{Env}}$, $\mathfrak{C})=$ \textsc{buildEnvirons}($I$, $P$, $C_\text{Inp}$, $C_\text{Env}$, $\mathfrak{C}$)
\STATE {}
\REPEAT \label{1st:line:loop1}
  \STATE {\emph{\textcolor{blue}{// First syncronize with input-fuzzing greybox fuzzer}}}
  \IF {new input seeds $I'$ found by vanilla greybox fuzzer}
    \STATE $(C_{\text{Inp}}$, $\mathfrak{C}$, $C_{\text{Env}})=$ \textsc{buildEnvirons}($I'$, $P$, $C_{\text{Inp}}$, $C_{\text{Env}}$, $\mathfrak{C}$)
  \ENDIF
  \STATE {}
  \STATE {\emph{\textcolor{blue}{// Then select and fuzz a random environment $S\in C_{\text{Env}}$}}}
  \STATE {Seed environment $S = $ \textsc{select}($C_{\text{Env}}$)}
  \STATE {Energy $e = $ \textsc{assignEnergy}($S$)}
  \STATE {}
  \FOR {\textbf{all} integers from $1$ to $e$}
    \STATE {\emph{\textcolor{blue}{// Mutate $S$ to generate $S'$ and execute $S'$}}}
    \STATE $S'=S(t',R') =$ \textsc{mutateEnviron}($S$)
    \STATE Preload \texttt{runtime} in \emph{fuzzing mode}
    \STATE Set resources $R'$ in \texttt{runtime}
    \STATE Execute program $P(t')$
    \STATE {}
    \STATE {\emph{\textcolor{blue}{// If $S'$ is interesting, update affected corpora}}}
    \IF{$P(t')$ with resources $R'$ increased coverage}
      \STATE {Add $S'$ to $C_{\text{Env}}$}
      \IF{$R'\neq R$}
        \STATE New resource $r'=R' \setminus R$
        \STATE Add $r'$ to $C_{\theta}\in \mathfrak{C}$ where $\theta$ is $r'$s resource identifier
      \ELSE
        \STATE Add $t'$ to $C_{\text{Inp}}$
      \ENDIF
    \ELSIF{$P(t')$ with resources $R'$ crashed}
      \STATE Add $S'$ to $C_{\text{\xmark}}$
    \ENDIF
  \ENDFOR
\UNTIL {$timeout$ reached or abort-signal} \label{1st:line:loopn}
\ENSURE {Crashing environments $C_{\text{\xmark}}$}
\end{algorithmic}
\end{algorithm}

\section{Coverage-Guided Mutation-based Environment Synthesis}\label{sec:mainalg}
\vspace{-0.05cm}
The overall procedure of coverage-guided mutation-based environment synthesis, \afc, is given in Algorithm~\ref{alg:fuzzing}. Like any other greybox fuzzer, \afc is started with a set of initial program inputs $I$ and the program $P$ (which can also represent a fuzz harness that executes fuzzer-generated program inputs on the harnessed part of the program). The main objective of \afc is to produce a (hopefully non-empty) set of bug-inducing program environments $C_{\text{\xmark}}$.

\textbf{Initialization} (Alg.\ref{alg:fuzzing}, L.1--9). \afc maintains three kinds of seed corpora. The input seed corpus $C_\text{Inp}$ is initialized with the initial program inputs $I$ while all other corpora are initially empty. For every initial input $t\in I$, \afc builds the corresponding environment seed, creates a resource-specific corpus for every accessed resources, and adds copies of the accessed resources as initial resource-specific seeds.

\begin{algorithm}
\caption{Function \textsc{buildEnvirons}}
\label{alg:buildenvs}
\begin{algorithmic}[1]
\REQUIRE {Input seeds $I$, Program $P$}
\REQUIRE {Input corpus $C_{\text{Inp}}$, Environment corpus $C_{\text{Env}}$}
\REQUIRE {Resource-specific corpora $\mathfrak{C}$}
\FOR {\textbf{all} $t \in I$}
  \STATE {Preload \texttt{runtime} in \emph{capture mode}}
  \STATE {Copied resources $R=P(t)$}
  \FOR {\textbf{all} $r \in R$}
    \IF {$C_r\not\in \mathfrak{C}$}
      \STATE {Resource-specific corpus $C_r = \{r\}$}
      \STATE {Resource-specific corpora $\mathfrak{C} = \mathfrak{C} \cup \{C_r\}$}
    \ENDIF
  \ENDFOR
  \STATE {Environment $S=S(t,R)$} 
  \STATE {Environment corpus $C_{\text{Env}} = C_{\text{Env}}\cup \{S\}$}
\ENDFOR
\ENSURE {Input corpus $C_{\text{Inp}}$, Environment corpus $C_{\text{Env}}$}
\ENSURE {Resource-specific corpora $\mathfrak{C}$}
\end{algorithmic}
\end{algorithm}
\textbf{Input-specific initial environments} (Alg.~\ref{alg:buildenvs}). A procedural overview of the environment construction for a set of program inputs $I$ is shown in Algorithm~\ref{alg:buildenvs}. For every program input $t\in I$, \afc executes the program $P$ on $t$ with the runtime preloaded and set to \emph{capture mode} (L.2). The runtime makes copies $R$ of every resource that is accessed during the execution of $t$ (L.3; cf. Sec.~\ref{sec:runtime}). For every copy $r\in R$, if there does not exist a resource-specific corpus $C_r$, \afc creates one with $r$ as initial seed, and adds $C_r$ to the set of resource-specific corpora $\mathfrak{C}$ (L.4--9). Finally, \afc builds the original environment $S$ for $t$ as $S(t,R)$ and adds it to the environment corpus $C_\text{Env}$ (L.10--11).

\textbf{Synchronization} (Alg.~\ref{alg:fuzzing} L.10--14). The first step of the main loop is to incorporate any new seeds $I'$ found by a vanilla fuzzer running concurrently (cf. Fig.~\ref{fig:arch}). \afc focuses on environment synthesis and delegates the generation of coverage-increasing program inputs to other greybox fuzzers. For every newly discovered program input $t\in I'$, \afc constructs the original environment for $t$ as discussed above.

\textbf{Selection and Power Schedule} (Alg.~\ref{alg:fuzzing} L.16--20). In order to steer the fuzzing process, environments can be subject to the same kind of optimizations that normal program inputs in others can be subject to. The power schedule determines the number $e$ of mutated environments that are generated from the selected environment $S$. In our implementation, the energy $e$ for $S$ depends on the number $|R|$ of resources accessed by $S=S(t,R)$. It is possible to implement the same kind of heuristics into the seed selection of power schedule, e.g., to prioritize environments that exercise rare program paths  \cite{aflfast} or branches \cite{fairfuzz}, or such environments that cover code that might be more buggy \cite{parmesan}.

\textbf{Mutation, execution \& corpus update} (Alg.~\ref{alg:fuzzing} L.21--39). The selected seed environment is mutated as discussed in \autoref{sec:fuzzer} for Algorithm~\ref{alg:mutate}. The runtime is configured to point to the resources $R'$ in the mutated seed environment $S'=S(t',R')$ and the input $t'$ is executed on program $P$ with the runtime preloaded in \emph{fuzzing mode}. If the execution of this alternative environment increased code coverage, \afc adds the coverage-inducing resource or input and the new environment to the corresponding corpora. If a crash was observable during execution, \afc adds the environment $S'$ to the set of crashing environments $C_{\text{\xmark}}$.

\section{Experimental Setup}
\subsection{Research Questions}
\noindent
Our experiments are guided by the following questions.
\begin{description}[leftmargin=1cm]
    \item[\textbf{RQ.1}] (Resources). \emph{Which environment resources do our bench\-mark programs access?} Qualitatively, we want to find out which part of the environment \afc can control (i.e., capture, mutate, and restore).
    \item[\textbf{RQ.2}] (Coverage Effectiveness). \emph{Compared to our baseline, how much more branch coverage does \afc achieve}? Quantitatively and qualitatively, we would like to understand which additional environment-dependent code is being covered when \afc synthesizes new program environments. We also investigate how effective the designed mutation operators are in terms of increasing code coverage? 
    \item[\textbf{RQ.3}] (Bug Finding Effectiveness). \emph{Compared to our baseline, how many more bugs does \afc expose}? We test our hypothesis that there exist environment-dependent bugs and that we can clearly attribute the discovered bugs to bug-inducing environments.
    \item[\textbf{RQ.4}] (Performance Overhead). \emph{How many test executions per second does \afc achieve compared to the baseline \aflnet}? This question is designed to reveal  the cost of the increased effectiveness.
    
\end{description}

\subsection{Fuzzer Implementation}

We chose the network-enabled greybox fuzzer \aflnet \cite{aflnet} as \emph{baseline}. We implemented our technique into \aflnet calling our tool \afc.
In addition to \aflnet, our extension \afc talks to our runtime and maintains a hierarchical queue structure. The runtime is implemented in the C programming language, as a standalone shared library.

During the fuzzing campaign, analogous to \aflnet, \afc outputs a result folder, which encompasses all the precious crafted seeds, along with fuzzer statistics and monitoring information. As there are three seed types in \afc, each one is stored accordingly.
\begin{enumerate}
    \item \emph{Input seed corpus}. In \aflnet, each input seed is stored as a file inside \emph{replayable-queue} directory on the file system. We have retained and used this structure.
    \item \emph{Resource-specific seed corpus}. We store these seeds in the \emph{Resources} directory. For each resource $C_{r}$, there is a sub-directory identified by the resource's identifier, in which the relevant resource-specific seeds will be stored as a file with unique identifier.
    \item \emph{Environment corpus}. We store these seeds in the \emph{system-level-seeds} directory. For reproducibility purposes, a specific structure is designed to store a particular program environment ($S(t,R)\in C_\text{Env}$) as a file on the file system. The file representing a program environment is a text file with the following format, which pinpoints the exact composition of that program environment:
    \begin{itemize}
    \item line 1: An integer $m$ representing the number of PUT's accessed resources.
    \item lines 2 to m+1: A resource identifier followed by the path to the corresponding resource-specific seed, stored on the file system. 
    \item line m+2: The path to the input seed, as is stored by \aflnet. 
\end{itemize}
 Crashing environments are stored in a separate sub-directory. It makes crash re-production more convenient.
\end{enumerate}

\subsection{Benchmarking}
\emph{Platform}. For our evaluation, we choose \textsc{ProFuzzBench} \cite{profuzzbench}, a mature benchmarking platform for network-enabled fuzzers that has been developed by the authors of \aflnet. All the experiments in this paper are conducted using this setup. The following benchmark programs have been made readily available in \textsc{ProFuzzBench}.

\emph{Benchmark}. \autoref{tab:benchmark} shows the seven selected real-world target programs as the subjects for our experiments. \texttt{Dcmtk} is an implementation for DICOM protocol, which is used for medical image exchange between hospitals. \texttt{Dnsmasq} is an implementation for DNS protocol. \texttt{Kamailio} is a SIP server for VoIP and realtime communications. \texttt{Live555} is a library for media streaming. \texttt{Openssh} is the premier connectivity tool for remote login with the SSH protocol. \texttt{OpenSSL} is a security-critical encryption library and finally, \texttt{PureFtpd} is a free FTP Server with a strong focus on software security.

\emph{Selection Criteria}. We selected our subject case studies based on the fact that they are well-known open-source network protocol implementations for widely used protocols, which additionally are 1) featuring resource access on the file system, as the main selection criteria, 2) widely used (as indicated in the Share column in \autoref{tab:benchmark} as the number of IPs using it, obtained from a search in Shodan \cite{shodan}) and 3) are used in security-critical industries such as medical in which any bug can be of serious consequences.

\begin{figure}
   
    \centering
    \footnotesize 
    \begin{tabular}{@{ \ }c@{ \ }|@{ \ }c@{ \ }c@{ \ }r@{ \ }r@{ \ }c@{ \ }}
        \textbf{Subject} & \textbf{Protocol} & \textbf{Commit} & \textbf{Size} & \textbf{Share} \\\hline
        DCMTK & DICOM & \texttt{7f8564c} & 952k LoC& 3k IPs \\
        Dnsmasq & DNS & \texttt{v2.73rc6 } & 35k LoC & 695k IPs\\
        Kamailio & SIP & \texttt{2648eb3}  & 1.1M LoC & 3k IPs\\
        Live555 & RTSP & \texttt{ceeb4f4} & 78k LoC& 12k IPs\\
        Openssh & SSH & \texttt{7cfea58} & 130k LoC & 39k IPs\\
        Openssl & SSL/TLS & \texttt{0437435} & 571k LoC& $>$10m IPs \\
        PureFtpd & FTP & \texttt{c21b45f}& 33k Loc & 1k IPs
    \end{tabular}
    \caption{Protocol implementations studied.} 
    \label{tab:benchmark}
\end{figure}

\subsection{Setup and Infrastructure}
All the experiments are conducted on a Ubuntu 18.04 machine with 32 vCPUs and 128GB of main memory, supplied by [blinded], a national research cloud that provides cloud computing services and tools to [blinded] researchers.
As elaborated in \autoref{sec:arch}, we ran an \afc instance in parallel with an \aflnet instance---with input synchronization enabled. For a fair comparison, we set up the baseline by running two synchronized instances of \aflnet.

We compared the baseline input-generating fuzzer and our proposed environment-generating fuzzer in terms of their branch coverage achieved, the number of bugs found, and the number of executions per second over fuzzing campaigns that lasted 24 hours. For each of tuple of the 2 fuzzers and 7 programs, we repeated each campaign 10 times to mitigate the impact of randomness.
\section{Experimental Results}
\subsection*{RQ1. Controllable Environment Resources}
Table \ref{tab:cat} shows the environment resources that \afc can control during the execution of the server programs. The types of resources are comprised of configuration files, databases, cryptographic keys and certificates, as well as rich content that is made available by the server. Most of the resources provide structured data. Even the configuration files follow different proprietary formats. This diversity of formats motivates the mutation-based approach of \afc. For 6 of 13 resources accessed by the server programs in our benchmark setup, we found bugs in the code that process these resources. This demonstrates the effectiveness of \afc to synthesize bug-inducing environments.

\begin{figure}

    \centering
    \footnotesize
    \begin{tabular}{@{}r@{ }|@{ }r@{ }|@{ }c@{ }|@{ \ }p{5cm}@{}}
        \textbf{Subject} & \textbf{Resource} & \textbf{Bug} & \textbf{Description} \\\hline
        DCMTK &  Configuration &\checkmark  & Server settings. Key-value text file.\\
              &   Database & & Index of all the stored images. Binary file.\\\hline
        Dnsmasq &  Configuration & & Server settings. Key-value text file.\\\hline
        Kamailio & Configuration &\checkmark & Server settings. Key-value text file.\\\hline
        Live555 &  Content & \checkmark & Multimedia files (mp3, wav, webm, mkv, webm, aac, ac3) to be streamed\\\hline
        Openssh  & Configuration & & Server settings. Key-value text file.\\\hline
        Openssl  & Crypto Key & \checkmark & Private key of server. PEM format.\\
                 & Cetificate & \checkmark & Server certificate. PEM format.\\
                 & Configuration & & Server settings. Key-value text file.\\\hline
        PureFtpd & Database &  \checkmark & Database used to keep user information, authentication, etc\\
                 & Content & & Files made available via the FTP server.\\
                   & Server Banner & & Message to show upon login. Text file.\\
                 & Configuration & & Server settings. Key-value text file.
    \end{tabular}
    \caption{Resources accessed during execution.} 
    \label{tab:cat}
\end{figure}

\begin{figure}[htp]\centering
\includegraphics[width=\columnwidth]{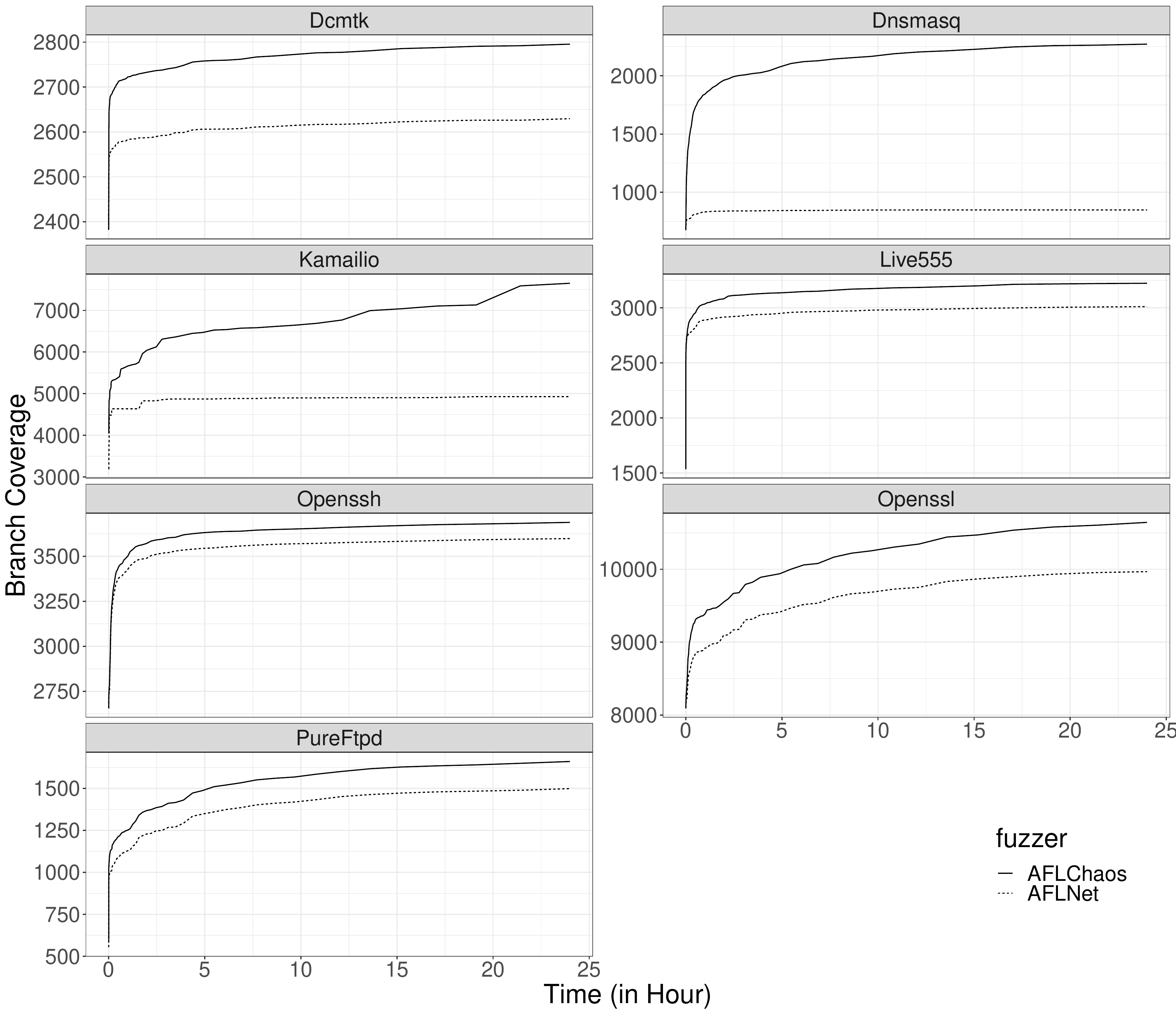}
\caption{Mean branch coverage over time for 10 runs of 24 hrs.}\label{fig:cov_plots_avg}
\end{figure}
\subsection*{RQ2. Coverage Effectiveness}
\textbf{Quantitative analysis}.
\autoref{fig:cov_plots_avg} shows the average branch coverage achieved over time across 10 campaigns of 24 hour for each fuzzer and subject. \autoref{tab:coverage} shows the effect size and statistical significance of the difference in effectiveness in terms of coverage for the 24 hour fuzzing campaigns.

In terms of coverage, \afc substantially outperformed \aflnet on all subjects. This indicates a significant improvement in terms of effectiveness when the program's environment is being made subject to fuzzing, as well. A change of the configuration for Dnsmasq and Kamailio brought about a 167.6\% and 55.2\% coverage improvement, respectively. For all other subjects, a coverage improvement above 5\% is considerable, given that most fuzzers that have been designed to improve efficiency essentially perform similarly~\cite{fuzzbench}.

\begin{figure}

    \centering\footnotesize
    \begin{tabular}{@{}c@{ \ }|@{ \ }r@{ \ }r@{ \ }r@{ \ }r@{ \ }|@{ \ }r@{}}
        \textbf{Subject} & \textbf{\afc} & \textbf{\aflnet} & \textbf{$\hat{A_{12}}$} & \textbf{p-value}& \textbf{Improvement}\\\hline
        DCMTK   & 2795.5 & 2629.5 & 1.00 & $<$0.001 & 6.3\%\\
        Dnsmasq & 2271.6 & 848.8  & 1.00 & $<$0.001 &167.6\% \\
        Kamailio & 7650.6 & 4929.0 & 0.96 & $<$0.001 &55.2\% \\
        Live555 & 3224.0 & 3011.6 & 1.00 & $<$0.001 &7.1\%\\
        Openssh & 3687.7 & 3597.7 & 1.00 & $<$0.001 &2.5\% \\
        Openssl & 10643.0 & 9967.3 & 1.00 & $<$0.001 & 6.8\% \\
        PureFtpd & 1659.5 & 1498.2 & 0.73 & 0.089 &10.8\%
    \end{tabular}
    \caption{Average branch coverage over 10 runs of 24 hours.\vspace{-0.2cm}} 
    \label{tab:coverage}
\end{figure}


\result{On our benchmark programs, on the median, \afc achieves 7\% more branch coverage than the baseline.}

\begin{figure}[]
    \centering\footnotesize
    \begin{subtable}[c]{0.5\textwidth}
     \begin{tabular}{@{}r@{ }|@{ }r@{ }|lr@{}}
    & \textbf{\#Files}     & \multicolumn{2}{c@{}}{\textbf{Coverage Increased Source Files} \hfill (\#Branches)}\\\hline
        Dcmtk   & 2211 &\checkmark dcmqrdb/libsrc/dcmqrcnf.cc & 83.4 \\
             & & \checkmark dcmqrdb/apps/dcmqrscp.cc &  19.6 \\
             & & \ \ \,ofstd/include/dcmtk/ofstd/oflist.h  & 11.5 \\
             & & \ \ \,ofstd/include/dcmtk/ofstd/ofmap.h  &  8.5 \\
             & & \checkmark ofstd/libsrc/ofcmdln.cc  & 8.0  \\
             & & \checkmark dcmdata/libsrc/dcdict.cc & 1.0\\
             & & \ \ \,dcmnet/libsrc/dccfenmp.cc& 1.0\\
             & & \ \ \,dcmnet/libsrc/dccfpcmp.cc & 1.0\\
             & & \ \ \,dcmnet/libsrc/dccfprmp.cc& 1.0\\
             & & \ \ \,dcmnet/libsrc/dccfrsmp.cc & 1.0\\
             & & \ \ \,dcmnet/libsrc/dccftsmp.cc& 1.0\\
             & & \checkmark oflog/libsrc/loglevel.cc & 1.0\\
             & & \ \ \,ofstd/libsrc/ofstring.cc& 1.0 \\
             \hline
         Live555 & 418 & \checkmark liveMedia/MatroskaFileParser.cpp  & 124.2 \\
             & & \checkmark liveMedia/MatroskaFile.cpp  &  33.5\\
             & & \checkmark liveMedia/VorbisAudioRTPSink.cpp  & 15.4 \\
             & & \checkmark liveMedia/MP3Internals.cpp  & 8.0 \\
             & & \checkmark liveMedia/ADTSAudioFileSource.cpp  & 7.5 \\
             & & \checkmark l$\sim$a/H264or5VideoStreamDiscreteFramer.cpp & 5.9 \\
             & & \checkmark l$\sim$a/OnDemandServerMediaSubsession.cpp   &  5.5\\
             & & \checkmark liveMedia/AC3AudioRTPSink.cpp  & 5.3 \\
             & & \checkmark liveMedia/AC3AudioStreamFramer.cpp & 4.5 \\
             & & \checkmark liveMedia/H264or5VideoRTPSink.cpp   & 3.8 \\
             & & \checkmark liveMedia/H264or5VideoStreamFramer.cpp  & 3.3 \\
             & & \checkmark liveMedia/H264VideoRTPSink.cpp & 3.0 \\
             & & \checkmark liveMedia/MP3StreamState.cpp & 2.5 \\
             & & \checkmark liveMedia/MultiFramedRTPSink.cpp & 2.4 \\
             & & \checkmark liveMedia/ServerMediaSession.cpp & 2.2 \\
             & & \checkmark liveMedia/MP3ADUdescriptor.cpp & 2.1 \\
             & & \checkmark testProgs/testOnDemandRTSPServer.cpp & 1.8 \\
             & & \ \ \,BasicUsageEnvironment/BasicHashTable.cpp &  1.0\\
             & & \checkmark liveMedia/MPEG4GenericRTPSink.cpp  &  1.0\\
             & & \checkmark liveMedia/RTCP.cpp & 1.0 \\
             \hline
    \end{tabular}
    \subcaption[]{\scriptsize Source files where \afc covered at least one new branch where directly environment-related files are indicated with a check mark (\checkmark). \#Files is the total number of source files. \#Branches is the average number of branches newly covered.}
    \end{subtable}\\[0.4cm]
    \begin{subtable}[c]{0.5\textwidth}\footnotesize\centering
    \begin{tabular}{@{}rrrrr}
    \textbf{Subject} & \textbf{\#Files} & \textbf{\#Cov. Incr.} & \textbf{\#Env.-Related} & \textbf{\#Branches}\\\hline
    Dnsmasq & 40   & 15 & 10 & {1322.0} \\
    Kamailio & 2802& 53 & 31 & {810.3}\\
    Openssh & 381  & 7 & 6 & {107.1}\\
    Openssl & 1610 & 63 & 44 & {822.7}\\
    \end{tabular}
    \subcaption[]{\scriptsize Aggregated results for the remaining subjects. For every program, we show the total number of source files (\#Files), the number of files where \afc covered at least one more branch (\#Cov. Incr.), the number of directly environment-related files (\#Env.-Related), and the total total number of branches only covered by \afc (\#Branches).}
    \end{subtable}
    \caption{Files where coverage was observed to increase.} 
    \label{tab:covq1}
\end{figure}
\textbf{Qualitative analysis}. We examined the code \emph{where} coverage increased to see if it depends on data provided by the accessed file resources. \autoref{tab:covq1} shows the (number of) source code files where at least one more branch was covered by \afc versus the total number of source files. It also shows the (number of) files which contain direct environment parsing/handling code.

In Figures~\ref{tab:covq1}.a and \ref{tab:covq1}.b, we observe that the increase in coverage is noticeable only in a small fraction of source files (e.g. 13/2211 for \texttt{DCMTK}, 20/418 for \texttt{Live555} and 53/2802 for \texttt{Kamailio}). It is interesting to note that most of these source files contain code that is directly related to the environment, e.g., because it opens the file and read the data (e.g. from a database), or because it parses and handles the content (e.g. different multimedia formats, PEM content or configuration). The source files with no checkmark often contain code that is indirectly related to the environment, mostly to implement data structures or basic object classes (e.g. hashmaps, lists, etc.) that capture the information extracted from these resources.

\result{The increase in coverage can be directly associated with environment-handling code.}

\begin{figure}

    \centering\footnotesize
    \begin{tabular}{@{}c@{ \ }|@{ \ }r@{ \ }@{ \ }r@{ \ }r@{ \ }|@{ \ }r@{}}
        \textbf{Subject} & \textbf{OP\#1} & \textbf{OP\#2} & \textbf{OP\#3} & \textbf{Total \#Seeds}\\\hline
        DCMTK & \textbf{365.7}& \underline{39.6} & 4.0 & 409.3\\
        Dnsmasq &  \textbf{874.3} & \underline{657.4} & 267.4 & 1799.0 \\
        Kamailio &  \textbf{915.6} &\underline{92.9} & 8.6 & 1017.0 \\
        Live555 &  \textbf{670.2} & 140.1 & \underline{236.3} & 1047.0\\
        Openssh &  \textbf{472.5} & \underline{404.0} & 15.7 & 892.2 \\
        Openssl &  \textbf{515.4} & \underline{106.1} & 9.1 & 630.6 \\
        PureFtpd & \textbf{164.5} & \underline{25.6} & 5.0&  195.1
    \end{tabular}
    \caption{Number of environment seeds that were added due to coverage-increasing changes originating from each mutation operator (average over 10 runs).} 
    \label{tab:ops}
\end{figure}

\textbf{Mutation operator effectiveness}. We developed three mutation operators for an environment seed  $S\in C_\text{Env}$. OP\#1 chooses a resource $r\in R$ at random, fuzzes the contents of the resource, and substitutes the chosen resource with the fuzzed version in the environment to generate an alternative environment $S(t, R')$. OP\#2 chooses a resource $r\in R$ at random and replaces the copy in the environment with another resource-specific seed chosen at random from the resource-specific corpus $C_r$ to generate an alternative environment $S(t, R')$. OP\#3 substitutes the input seed $t$ in the environment with another input seed $t'$ chosen at random from the input seed corpus $C_I$ to generate an alternative environment $S(t', R')$.
In preliminary experiments, we found OP\#1 to be most effective. To make better use of the fuzzing time, we increased the probability of selecting OP\#1 to 80\% while adjusting the probabilities for OP\#2 and OP\#3 to 10\%, respectively. The following result should be interpreted under this adjustment.

\autoref{tab:ops} shows for each mutation operator OP the number of environment seeds that were added as coverage-increasing environments as a consequence of applying OP. We can see that the contribution of OP\#1 to generate coverage-increasing environments is 2.71 and 7.28 times better than OP\#2 and OP\#3, respectively. In fact, we find that all three operators are necessary. Without OP\#2 and OP\#3, our approach of mutating just one resource at a time and not mutating input (i.e., receiving input seeds via syncing with a parallel AFLNet) would \emph{not} work. This is because the coverage of some code may depend on the content of more than just one resource and input seed as well. Towards this end, OP\#2 and OP\#3 provide mixing and matching of all the resources' content based on already found coverage-increasing resource seeds---incrementally constructing system-level seeds with several mutated resources and inputs.

\result{Mutation operator OP\#1 which chooses a random resource and fuzzes it appears to be most effective compared to the operators that switch resource (OP\#2) or input (OP\#3).}

Overall, these coverage results and analysis are highly promising regarding the ability of \afc in reaching and covering new lines of the mostly environment-related code that otherwise would remain uncovered.

\subsection*{RQ3. Bug Finding Effectiveness}
\begin{figure}[]
    \centering
    \footnotesize
    \begin{tabular}{@{}l@{  }|@{ }l@{ }r@{ }|@{ \ }l@{ \ }|@{ \ }l@{}}
        \textbf{Subject} & \multicolumn{2}{l}{\textbf{Bug Type} and [Bug Report]}& \textbf{24h Runs} & \textbf{48h Runs} \\\hline
        \multirow{4}{*}{DCMTK} & \textbf{Stack Overflow} &\cite{dcmtk_bug1} & 10h35m (01/10) & 26h51m (1/3)\\
         & \textbf{Buffer Overread} &\cite{dcmtk_bug1} & 00h33m  (10/10) & 00h20m (3/3)\\
         & \textbf{Buffer Overwrite} &\cite{dcmtk_bug1} & 05h13m  (04/10) & 45h06m (1/3)\\\hline
        \multirow{3}{*}{Kamailio} & \textbf{Buffer Overflow} &\cite{kamailio_bug1} & 06h55m  (03/10) & 10h49m (1/3)\\
         & \textbf{NULL Pointer 1} &\cite{kamailio_bug1} & 11h41m (04/10) &  06h48m (3/3)\\
         & \textbf{NULL Pointer 2} &\cite{kamailio_bug1} & 02h17m  (10/10) &  00h17m (3/3)\\\hline
        \multirow{3}{*}{Live555} &  \textbf{Heap Use After Free 1} &\cite{live_bug1}  & 03h54m  (09/10) & 04h09m (3/3)\\
        &  \textbf{Heap Use After Free 2} &\cite{live_bug2} & ---:---h (00/10) & 34h50m (1/3)\\
        & \textbf{NULL Pointer} &\cite{live_bug3} & 00h04m (10/10) & 00h02m (3/3)\\\hline
        \multirow{1}{*}{Openssl} & \textbf{Heap Use After Free} &\cite{openssl_bug} & ---:---h (00/10) & 12h08m (1/3)\\\hline
        \multirow{3}{*}{PureFtpd} & \textbf{Buffer Overread} &\cite{pureftpd_bug1} & 00h02m (10/10) & 00h03m (3/3) \\
         & \textbf{Heap Overflow} &\cite{pureftpd_bug3} & 00h22m (2/10) &01h41m (2/3) \\
         & \textbf{Division by Zero} &\cite{pureftpd_bug2} & 00h02m (10/10) & 00h02m (3/3) \\\hline
    \end{tabular}
    \vspace{0.2cm}
    \caption{Bugs found by \afc, time to discover each bug, and the number of trials in which the bug is found. If a bug is found in more than one campaign, the average time to discovery is reported. \aflnet found none of those.}
    \label{tab:bugtime}
    
\end{figure}

\textbf{Quantitative analysis}. \autoref{tab:bugtime} shows the results in terms of bug finding. \afc found \emph{11 bugs} in ten campaigns of 24 hours across the seven benchmark programs. All of those bugs were discovered in environment handling code, such that the baseline \aflnet was unable to find them. Five bugs were found in under one hour, on the average while the remaining bugs were found in between 2 and 12 hours. Five bugs were found consistently across all ten campaigns while the remaining bugs were found in between 1 to 9 campaigns.
Given the large variance of the time to exposure, we decided to run another three campaigns of 48 hours. This revealed two more bugs, increasing the total to 13 bugs found. Apart from the bugs in PureFTPd, all discovered bugs have been fixed by the developers. PureFtpd does not provide a public issue tracker and maintainers did not react to our emails. Our reports on an (unmaintained) fork remain unanswered \cite{pureftpd_bug1}-\cite{pureftpd_bug3}. Dcmtk and Kamailio bugs have been acknowledged and fixed by the developers \cite{dcmtk_bug1}, \cite{kamailio_bug1}. Live555 bugs were reported \cite{live_bug1}-\cite{live_bug3} and have been fixed (i.e., we cannot reproduce the bugs on the most recent version). We could not identify the specific bug fixing commit due to unavailability of the intermediate versions. The OpenSSL bug has been fixed \cite{openssl_bug}.

\result{\afc found 13 previously undiscovered bugs in five of the seven server programs that \aflnet did not find.}

\textbf{DCMTK}.
\afc found a critical buffer overwrite, a buffer overread, and a stack overflow bug in the DCMTK server.
The crash and fix locations for all bugs are in \texttt{dcmqrcnf.cc} which parses the server's configuration file.
Of course, all bugs are important and should be removed, but for this bug there is also a security risk (even if the attack vector is unlikely). If a malicious user can modify the configuration file, e.g., by social engineering (e.g., via a tutorial online) or by accessing the server machine as an unprivileged user, they can potentially kill the server or worse, gain root privileges by privilege escalation. DCMTK implements the DICOM protocol, the international standard to transmit, store, retrieve medical imaging information. Privileged access would reveal sensitive medical data.


\textbf{Kamailio}. 
\afc finds three bugs in Kamailio; one global buffer overflow in the \texttt{core/route\_struct.c} file and two null pointer dereferences in the \texttt{core/rvalue.147} and \texttt{core/sr\_module.c}. 
The root causes of all the bugs are found in the source code that parses the server's configuration file (e.g. the \texttt{yyparse} file \texttt{core/cfg.y}) and also in the codes that apply the settings. All three bugs are resolved in a single bug-fixing commit \cite{kamailio_bug1}, which makes the program exit once the first faulty setting is detected in the configuration file (instead of issuing a bad-configuration warning) and prevents the crash-producing setting from being applied. 
In addition to improved software correctness, the discovery and patching of these bugs also reduced the security risks. Kamailio is a VoIP and Instant Messaging protocol, such that privileged access might reveal private user data.


\textbf{Live555}.
\afc found two critical heap-use-after-free bugs and a null pointer dereference in Live555.
The root causes for all bugs are located in the files handling the multimedia files to be streamed upon a remote user's request, i.e., \texttt{MatroskaFile.cpp}, \texttt{MPEG1or2Demux.cpp}, and \texttt{ADTSAudioFileServerMediaSubsession.cpp} which parse Matroska, MPEG and ADTS multimedia files, respectively.
If a malicious, unprivileged user can modify the multimedia files to be streamed, e.g., by uploading and the user requests the multimedia file for streaming, they can potentially kill the server or worse, gain root privileges by privilege escalation.

\textbf{OpenSSL}.
\afc found a critical heap-use-after-free in OpenSSL.
The root cause for this bug is located in code that parses (Privacy Enhanced Mail) PEM certificates and keys, i.e., \texttt{pem\_pkey.c}.
If an unprivileged user registers a malicious certificate or a malicious key provided in the PEM format, the heap-user-after-free may be exploited to gain priviliged access to the machine running the server. OpenSSL is a library used to encrypt communication such that an exploit would likely reveal sensitive information.

\textbf{PureFTPd}.
\afc found a critical heap overflow, a buffer overread, and a division by zero bug in PureFTPd.
The root causes for these bugs are located in the code that parses and handles the PureDB database.
The ability to write to the PureDB database as an unprivileged user might allow a malicious user to exploit one of these bugs in PureFTPd for denial of service or privilege escalation attacks. For PureFTPd, we reported the bugs but have not heard back.

\result{All bugs that \afc discovered are induced by environment resources that are not related to server inputs (i.e., user requests). The root causes of those environment-induced bugs are located in source code that is handling configuration files, databases, certificates and keys, as well as multimedia files intended for streaming by a streaming protocol. If a malicious, unprivileged user can modify these environment resources of the corresponding privileged process, they may launch a privilege escalation attack via those bugs identified as critical.}

\subsection{Performance analysis}
The increased effectiveness comes at a cost, i.e., the additional time spent intercepting syscalls and copying resources. To assess this cost, we measure the number of executions per second for both fuzzers across all programs. As \afc is implemented into \aflnet, we can attribute the difference in performance to fuzzing the program environment instead of the program inputs.

\begin{figure}
    \centering
    \includegraphics[width=0.9\columnwidth]{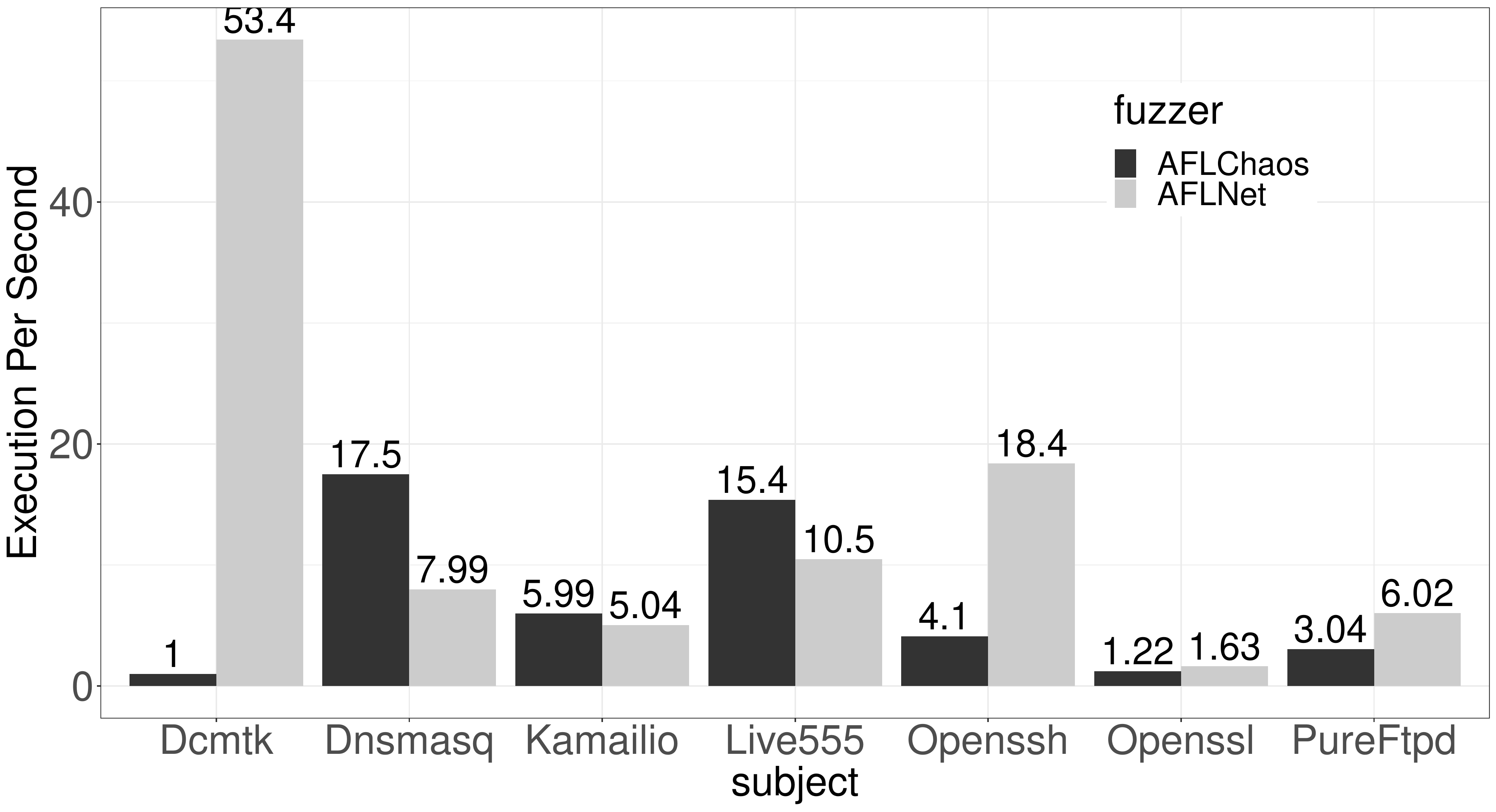}
    \caption{Performance differences between \afc and its baseline \aflnet in terms of the average number of executions per second.}
    \label{fig:eps}
\end{figure}

Figure \ref{fig:eps} shows the results of our performance analysis. For four out of seven programs, \aflnet generates substantially more executions per second than \afc. Specifically, \afc generates 25\%, 49\%, 78\%, and 98\% \emph{less} executions per second than \aflnet for openssl, pureftpd, openssh, and dcmtk, respectively. Upon closer investigation, we found that the corrupted resources sometimes led to unresponsiveness (a.k.a. ``hangs"), which requires the fuzzer to wait until the connection times out. These environment-induced timeouts affect the performance dramatically, specifically for dcmtk and openssh. This means that fuzzing the environment as implemented in \afc can indeed cause a substantial performance overhead. However, we believe (i)~that this additional overhead is reasonable given the substantial increase in effectiveness, and (ii)~that there are many opportunities to improve the engineering of our research prototype \afc (which we make publicly available).

For the remaining three out of seven programs, \afc generates more executions per second than \aflnet. Specifically, \afc generates 18\%, 46\%, and 119\% \emph{more} executions per second than \aflnet for kamalio, live555, and dnsmasq, respectively.
This speedup was a unexpected, at first. Upon closer investigation, we found that the corrupted resources sometimes led to early termination. For kamalio and dnsmasq, an execution fails quickly if the configuration file is corrupted. For Live555, an execution fails before the streaming commences if the parsing of the multimedia file fails.

\result{Given the increase in coverage and bug finding effectiveness, the performance cost is adequate. In some cases, the performance of \afc versus \aflnet even increases due to early termination of an execution with a corrupted environment resource.}

\section{Case Studies}\label{sec:motiv}
To explore the impact of the environment-induced program bugs that \afc discovered, we present the results for one program as a case study. As program, we chose DCMTK which implements the Digital Imaging and Communications in Medicine (DICOM) \cite{dicom} protocol. The DICOM protocol is the standard for the communication and management of medical imaging information and related data. For instance, DICOM is used in \texttt{NVIDIA Clara}\footnote{\scriptsize
\url{https://docs.nvidia.com/clara/deploy/RunningReferencePipeline.html\#install-dcmtk}} a container-based, cloud-native development and deployment framework for multi-AI, multi-domain workflows in smart hospitals and \texttt{Orthanc}\footnote{\url{https://book.orthanc-server.com/faq/dicom-tls.html\#using-dcmtk}} a free and open-source, lightweight DICOM server for medical imaging from Belgium.

\textbf{Criticality}. In the health care domain, the DCMTK library is a widely used software component. DCMTK deals with sensitive patient data. So, it is of critical importance to test whether there are any threats to confidentiality or availability. In fact, previously several CVEs have been reported against DCMTK,\footnote{\url{https://cve.mitre.org/cgi-bin/cvekey.cgi?keyword=dcmtk}} including a privilege escalation.\footnote{\url{http://hmarco.org/bugs/dcmtk-3.6.1-privilege-escalation.html}} All of these critical bugs have been patched promptly.

\textbf{Environment}. DCMTK is a privileged process:\footnote{\scriptsize\url{https://github.com/DCMTK/DCMTK/blob/master/dcmqrdb/docs/dcmqrcnf.txt\#L62}}. In our fuzzing set up, an execution corresponds to starting the server and processing the package that is constructed by the fuzze by mutation of recorded packages \cite{aflnet}. During execution, the DCMTK server accesses several environment resources (cf. \autoref{tab:cat}), including the following files
\begin{enumerate}
    \item \texttt{dcmqrscp.cfg} (Server settings)
    \item \texttt{./ACME\_STORE/index.dat} (Database)
    \item \texttt{/etc/hosts} (Database)
\end{enumerate}
Since there is no reason to assume that \texttt{/etc/hosts} can be corrupted intentionally or unintentionally, we blocklisted this file during our experiments. This file is Unix-specific, provides a mapping of some hostnames to IP addresses, and can only be accessed with root privileges.

We ran \afc on DCMTK using the existing harnessing available in ProFuzzbench \cite{profuzzbench} for one day and found three bugs (cf. \autoref{tab:bugtime}). All of these bugs can only be found by changes to the configuration file \texttt{dcmqrscp.cfg}.
Our \emph{threat model} assumes that an unprivileged attacker on the local machine  can modify the environment (i.e., external state) of the protocol implementation---in this case the configuration file. By modifying the unprivileged environment of a privileged process, the attacker can leverage the privilege of the protocol process to launch an arbitrary code execution attack on the host machine. For the three bugs we found in DCMTK specifically, there may the following consequences.

\begin{itemize}
  \item \emph{Privilege Escalation}. One bug is a buffer overwrite. Due to an unchecked \texttt{memcpy}, it can be exploited to write to otherwise unwritable memory \footnote{\scriptsize\url{https://github.com/DCMTK/dcmtk/blob/master/dcmqrdb/libsrc/dcmqrcnf.cc\#L706}} Since the protocol implementation is started as a privileged process, privilege escalation is a viable attack. The second bug is a stack overflow, due to an unchecked \texttt{strcmp} \footnote{\scriptsize\url{https://github.com/DCMTK/dcmtk/blob/7f8564c/dcmqrdb/libsrc/dcmqrcnf.cc\#L691}}, it might also be useful to gain unauthorized access to the system.
  \item \emph{Leaking Sensitive Data}. One of the bugs is a buffer overread. Due to a \texttt{strcmp} being executed on a non-null-terminated string, the attacker can read memory, they are not supposed to read.\footnote{\scriptsize\url{https://github.com/DCMTK/dcmtk/blob/master/dcmqrdb/libsrc/dcmqrcnf.cc\#L686}} Medical images are particularly sensitive data. Even a buffer overread can leak sensitive patient data.
\end{itemize}
These bugs \emph{cannot} be found by \aflnet which fuzzes only the input space of DCMTK (i.e., the packages it processes).

\section{Related work}
\textbf{Snapshotting}. The idea of capturing and restoring the runtime state of a machine, including the programs that are running on it dates back about thirty years ago; but it wasn't until the early 2000's that the required hardware-assistance was built into the first processors \cite{intel}. The system that runs virtual machines on a host machine is called hypervisor. To improve the efficiency of fuzzing, particularly of stateful systems, researchers have developed fuzzers using the principle of snapshot-restore upon such hypervisors \cite{Nyx2022,nyxnet}. However, existing works do not actually modify the environment within which the program-under-test runs. Since fuzzing the environment itself might yield unrealistic environments or may not impact the program behavior, we propose to fuzz the \emph{interaction} with the program's environment instead.

\textbf{Domain-specific approaches}. There exist several domain-specific approaches to fuzz non-traditional program ``inputs". For instance, there are domain-specific works on fuzzing a program's configuration \cite{Confuzzing2010,CEI2021,fuzzingbook2022:ConfigurationFuzzer}, databases \cite{db2020,db2020Ri,db2020Ma,db2008,csmith,emi}, the network \cite{networkfuzzing2021,aflnet,pulsar,Nyx2022}, et cetera. For instance, in the Fuzzing Book \cite{fuzzingbook2022:ConfigurationFuzzer}, the importance of testing all possible configuration options passed to the program is addressed by proposing a way of automatically inferring configuration options. The program's configuration can then be tested using combinatorial  testing \cite{cit}. However, all previous approaches focus on domain-specific improvements.
There are programs that take non-traditional inputs. For instance, database management systems (DBMS) take databases as input; compilers take program source code as inputs. In order to test these systems automatically domain-specific approaches have been develop that generate these non-traditional inputs (i.e., databases and programs) \cite{db2020,db2020Ri,db2020Ma,db2008,csmith,emi}. However, these domain-specific approaches cannot be applied to test general programs, e.g., that use databases.
Our work can be viewed as a general unification by fuzzing, without discrimination and in a coverage-guided manner, all environment resources that the program-under-test accesses.

\textbf{Chaos engineering}. The key idea behind chaos engineering is to arbitrarily modify a running system and observe the system's ability (i.e., robustness) to handle those faults \cite{chaos}.  Basiri et al \cite{chaos} discuss Chaos engineering as an experimental discipline in which the software is viewed as a set of processes, each may fail arbitrarily. For instance, a memory allocation (malloc) may return that no more memory can be allocated, a disk read or write may fail simulating a full disk or absent file, and so on. As real failures are hard to reproduce, fake failures are injected automatically in production to inspect how the system reacts and to find bugs.
In contrast to chaos engineering, we do not suggest to modify a running system \emph{in production}. Instead, we propose coverage-guided mutations to the program's interactions with the environment.

In the concrete case of program testing, Jiang et al. \cite{Zu2020} proposed a chaos-inspired fault-injection to find bugs in error-handling code \cite{Zu2020}. The authors introduce a context-sensitive software fault injection (SFI) approach, in which the error handling code of system calls is fuzzed by injecting arbitrary faults in different contexts into the program and find hard-to-find bugs in the error handling code. In contrast to Jiang et al., we propose the coverage-guided mutation of environment resources rather than simple fault injection (e.g., where malloc returns 0). While Jiang et al.'s objective is to find bugs in error-handling code specifically, our objective is to find bugs in environment-handling code more generally.

\section{Conclusion}
The contribution of this paper is to extend the fuzzer's search space beyond the traditional input space to include all environment resources that the program interacts with. These accessed environment resources fully capture the pertinent external state of the program. Our approach does not take the abstract view of the program as an isolated process that receives input and produces output, but more realistically, it considers the context within which the program is being executed. Our work features a fine-grained approach to the external state, in which each external element can be identified and handled separately. We have defined the novel definition of system-level seeds which can incorporate the external elements along with the input seed. This enables pinpointing the reason for found crashes as well as straight-forward crash reproduction. In our prototype, we focus on files as an important element of the external environment, given that in Linux, everything is considered a file. Our experimental results on seven widely used protocol implementations showed outstanding increase in code coverage for \afc in comparison with \aflnet as the base and resulted in discovering 13 bugs in these widely-used and well-fuzzed programs.
\section{Data Availability}
\textbf{Open Science}. To facilitate open science and reproducibility, we have published our research prototype, our experimental setup, all data, and our analysis:\\
{%
\url{https://anonymous.4open.science/r/AFLChaos\_Artifact-024B}
}

\bibliographystyle{ACM-Reference-Format}
\bibliography{references}  

\end{document}